  \documentclass{icrc2009}

  \usepackage{graphicx}   
  \usepackage[font=footnotesize]{subfig} 
  \usepackage{fixltx2e}
  \usepackage{url}

  \newcommand{\shorttitle}[1]%
  {\markboth{Proceedings of the 31\MakeLowercase{$^{st}$} ICRC, {\L}\'{o}d\'{z} 2009}{#1} }
  \newcommand{\etal}{\MakeLowercase{\textit{et al. }}} 

\usepackage[usenames,dvipsnames]{color}
\usepackage{hyperref}
\definecolor{orange}{rgb}{1,0.5,0}
\hypersetup{
    colorlinks,%
    citecolor=RoyalBlue,%
    filecolor=RedViolet,%
    linkcolor=ProcessBlue,%
    urlcolor=RedViolet
}

\begin{document}
  \title{Operating Water Cherenkov Detectors in high altitude sites for the
Large Aperture GRB Observatory}

\author{ \vbox{ D.~Allard$^a$,
C.~Alvarez$^b$,
H.~Asorey$^c$,
H.~Barros$^d$,
X.~Bertou$^c$,
M.~Castillo$^e$,
J.M.~Chirinos$^f$,
A.~De Castro$^d$,
S.~Flores$^g$,
J.~Gonzalez$^h$,
M.~Gomez Berisso$^c$,
J.~Grajales$^e$,
C.~Guada$^i$,
W.R.~Guevara Day$^j$,
J.~Ishitsuka$^g$,
J.A.~L\'opez$^k$,
O.~Mart\'inez$^e$,
A.~Melfo$^i$,
E.~Meza$^l$,
P.~Miranda Loza$^m$,
E.~Moreno Barbosa$^e$,
C.~Murrugarra$^d$,
L.A.~N\'u\~nez$^i$,
L.J.~Otiniano Ormachea$^j$,
G.~Perez$^n$,
Y.~Perez$^i$,
E.~Ponce$^e$,
J.~Quispe$^m$,
C.~Quintero$^i$,
H.~Rivera$^m$,
M.~Rosales$^i$,
A.C.~Rovero$^o$,
O.~Saavedra$^p$,
\underline{H.~Salazar}$^{e,1}$,
J.C.~Tello$^d$,
R.~Ticona Peralda$^m$,
E.~Varela$^e$,
A.~Velarde$^m$,
L.~Villasenor$^q$,
D.~Wahl$^g$,
M.A.~Zamalloa$^r$
 (LAGO Collaboration)}\\
{$^a$}APC, CNRS et Universit\'e Paris 7. France\\
{$^b$}Universidad Autonoma de Chiapas, UNACH.  M\'exico \\
{$^c$}Centro At\'omico Bariloche, Instituto Balseiro.  Argentina \\
{$^d$}Laboratorio de F\'isica Nuclear, Universidad Sim\'on Bol\'ivar, Caracas. Venezuela \\
{$^e$}Facultad de Ciencias F\'isico-Matem\'aticas de la BUAP.  M\'exico \\
{$^f$}Michigan Technological University. USA\\
{$^g$}Instituto Geofisico del Per\'u, IGP. Lima - Per\'u\\
{$^h$}Universidad de Granada. Spain\\
{$^i$}Universidad de Los Andes, ULA. M\'erida - Venezuela\\
{$^j$}Comisi\'on Nacional de Investigaci\'on y Desarrollo Aeroespacial, CONIDA. San Isidro Lima - Per\'u\\
{$^k$}Universidad Central de Venezuela, Facultad de Ciencias, Departamento de F\'isica. Venezuela\\
{$^l$}Universidad Nacional de Ingenieria, UNI. Lima 25 - Per\'u\\
{$^m$}Instituto de Investigaciones F\'isicas, UMSA. Bolivia \\
{$^n$}Universidad Polit\'ecnica de Pachuca.  M\'exico \\
{$^o$}Instituto de Astronom\'ia y F\'isica del Espacio. Argentina \\
{$^p$}Dipartimento di Fisica Generale and INFN, Torino. Italy\\
{$^q$}Instituto de F\'isica y Matem\'aticas, Universidad de Michoac\'an. M\'exico\\
{$^r$}Universidad Nacional San Antonio Abad del Cusco. Per\'u
}

  \shorttitle{H. SALAZAR \etal LAGO Sites}
  \maketitle

  \begin{abstract}
  Water Cherenkov Detectors (WCD) are efficient detectors for detecting
GRBs in the 10 GeV - 1 TeV energy range using the single particle
technique, given their sensitivity to low energy secondary photons
produced by high energy photons when cascading in the atmosphere. The
Large Aperture GRB Observatory (LAGO) operates arrays of WCD in high
altitude sites (above 4500\,m\,a.s.l.) in Bolivia, Mexico and Venezuela,
with planned extension to Peru. Details on the operation and stability
of these WCD in remote sites with high background rates of particles
will be detailed, and compared to simulations. Specific issues due to
operation at high altitude, atmospheric effects and solar activity,
as well as possible hardware enhancements will also be presented.
\footnotetext[1]{presenting and corresponding author, \href{mailto:hsalazar@fcfm.buap.mx}{hsalazar@fcfm.buap.mx}}
  \end{abstract}


  \section{Introduction}
\begin{figure}[!ht]
    \centering
    \includegraphics[width=0.5\textwidth]{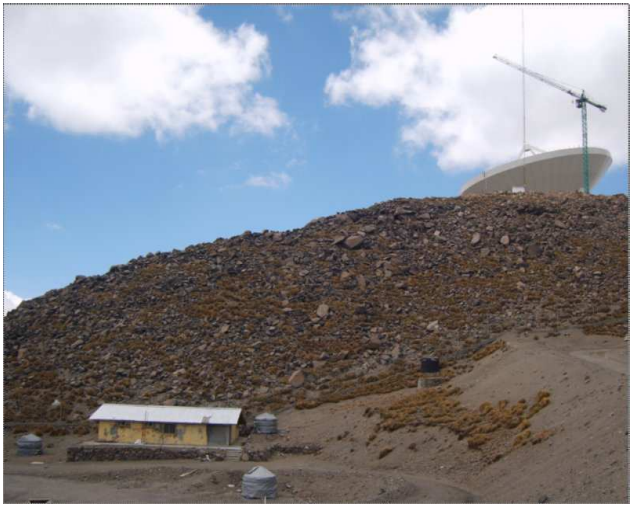}
    \caption{LAGO-Sierra Negra site Mexico at 4550\,m\,a.s.l.}
    \label{lago:mex}
\end{figure}

\begin{figure}[!ht]
    \centering
    \includegraphics[width=0.5\textwidth]{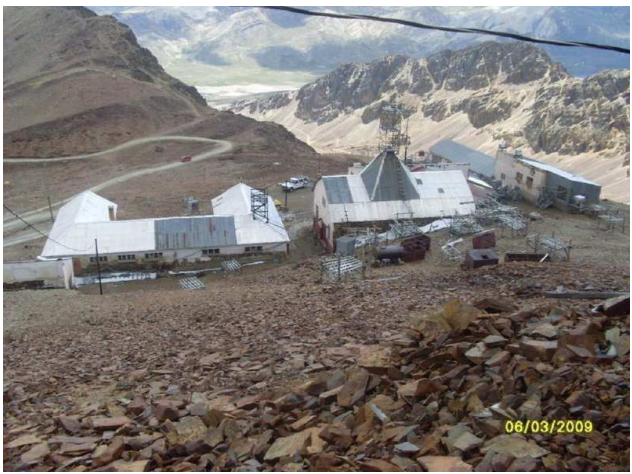}
    \caption{LAGO Chacaltaya site at 5280\,m\,a.s.l.}
    \label{lago:boli1}
\end{figure}

\begin{figure}[!ht]
    \centering
    \includegraphics[width=0.5\textwidth]{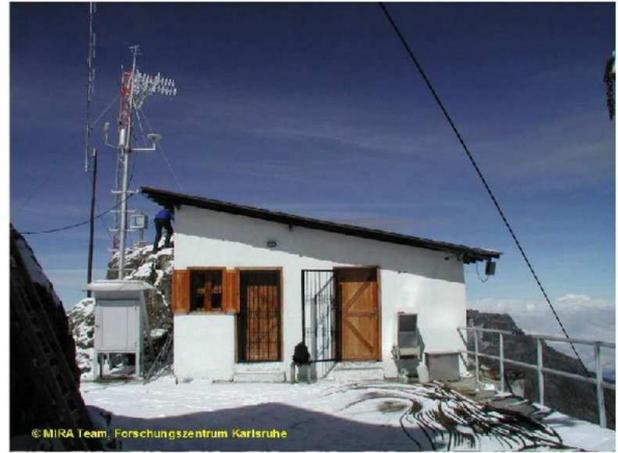}
    \caption{LAGO Merida site at 4780\,m\,a.s.l.}
    \label{lago:vene}
\end{figure}

  Since photons coming from GRBs can not penetrate easily the atmosphere, it is necessary to use
 satellites to detect them. However, as the photon energies increase, the photon flux decreases
 as a power law. Therefore, in order to detect  small fluxes of gamma radiation or high energy 
photons in the range of GeV to TeV is  necessary to construct  more sensitive detectors with
 larger areas. Satellites with large collecting areas  become impractical due to their cost.
 However, with inexpensive ground-based experiments of large area, it is possible to detect 
the relativistic secondary particles induced by the interaction of GeV or TeV gamma-ray 
photons with the molecules of the upper atmosphere. Water Cherenkov Detectors (WCD) are efficient
 detectors 
for detecting GRBs in the 10 GeV - 1 TeV energy range using the single particle technique, 
given their sensitivity
 to low energy secondary photons produced by high energy photons when cascading in the
 atmosphere.\\
Currently or in the recent past, a handful of ground-based experiments around the world 
 are searching 
GRBs: Chacaltaya at 5200\,m\,a.s.l. in Bolivia (INCA); Argo at 4300\,m\,a.s.l. in Tibet;
 Milagro at 
2650\,m\,a.s.l. in New Mexico; the Pierre Auger Observatory at 1400\,m\,a.s.l. in Malarg\"ue,
 Argentina  and  LAGO (with several sites).  Of all these experiments only the prototype
 of Milagro called Milagrito has reported the possible detection of signals associated to 
a GRB, GRB 970417. Milagro is the largest area (60\,m$\times$80\,m) water Cherenkov detector
 capable of continuously monitoring the sky at energies between 250 GeV and 50 TeV. Although
 designed to study ultra high energy cosmic rays, the Pierre Auger Observatory is also a 
competitive high energy GRB ground-based detector due to its large area and the good 
sensitivity to photons of its water Cherenkov detectors . The Large Aperture GRB Observatory 
(LAGO) operates arrays of WCD in high altitude sites (above 4500\,m\,a.s.l.) in Mexico, Bolivia
 and Venezuela (see figs.~\ref{lago:mex}, \ref{lago:boli1}, \ref{lago:vene}), with planned 
extension to Colombia, Guatemala, the Himalaya and Peru.\\
\\
\\
\\
Details on the operation and stability of these WCD in remote sites with high background rates
 of particles will be detailed. Specific issues due to operation at high altitude, atmospheric
 effects, as well as possible hardware enhancements will also be presented.\\
  Two basic conditions are needed to achieve a more sensitive water Cerenkov detector: a higher
 site and a larger area detector. LAGO is looking for the first option increasing the altitude of 
the detectors and spreading their detectors in several latitudes.\\
 Further improvements can be made in the design of the detector in order to reach full autonomy,
 low energy consumption, easiness in deployment and remote monitoring system.

  \section{The LAGO sites}
  The LAGO project aims at observing GRBs by the single particle technique using water Cherenkov
 detectors (WCD). It would consist of various sites of large efficiency to GRB detection for their
 altitude. Suitable sites above 4500 metres with support infrastructure, namely an access road, 
electricity and Internet are hard to find. However LAGO requires a very small flat area 
and should not need human operators in the site. \\

\begin{figure}[!ht]
    \centering
    \includegraphics[width=0.5\textwidth]{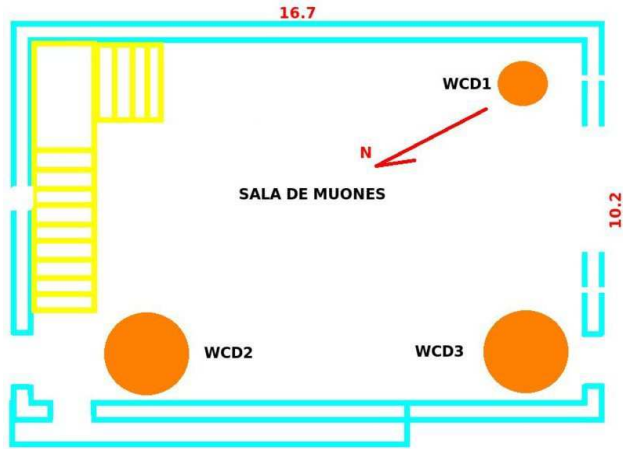}
    \caption{LAGO Chacaltaya Layout.}
    \label{lago:layout}
  \end{figure}

The Sierra Negra volcano is the site of the Large Millimetre Telescope/Gran 
Telescopio Milimetrico (LMT/GTM). The development of the LMT/GTM site 
started in 1997 with the
 construction of the access road, followed with the installation of
 a power line and an optical fibre link to the Internet, both currently functional.\\
  Sierra Negra is inside the Parque Nacional Pico de Orizaba, named after Pico
 de Orizaba, the highest mountain in Mexico with 5610\,m\,a.s.l. The National Park
 comprises both volcanoes, whose summits are separated by 7 km. Pico de Orizaba
 is a potential site for a second stage of LAGO project.Sierra Negra at 4550\,m\,a.s.l. is
 first LAGO  site with water Cherenkov detectors working (since early 2007). 
Currently 3$\times$4\,m$^{2}$ area Cherenkov detectors in a 30\,m triangular array
 are taking data at this site. \\
Monte Chacaltaya at 5270\,m\,a.s.l. is the highest and older observatory in the
 world having the Pion discovery has one of the most important results. Currently
 3 cherenkov detectors are  taking data at this site: two of them of 4\,m$^{2}$ 
area and the third one of 2\,m$^{2}$ area. They are positioned in an 15\,m $\times$
 10\,m rectangular array. Measurement of the atmospheric pressure and temperature,
 as well as two neutron monitors are included in the site   see fig.\,\ref{lago:layout}.\\
 \\

The LAGO Pico Espejo station is situated at the last station of the Merida Cable
 Car (Teleferico de Merida), at 4765 metres, overlooking the city of Merida. The 
data acquiring system will be situated in the Humboldt Research Station, a facility
 provided by Universidad de Los Andes.  As a part of the international LAGO collaboration,
 three water Cherenkov detector are being installed at Pico Espejo, in the Sierra Nevada
 National Park, Merida, Venezuela. The project is a collaboration between Universidad de
 Los Andes at Merida, Universidad Simon Bolivar, at Caracas, and the LAGO project.\\
Recently Peru joined the LAGO collaboration and its first prototype is under construction.
 High altitude site near Lima and in the Cuzco region are under consideration for the Peruvian LAGO site.\\

 \section{Operation and monitoring}
The operation of the running detectors at Chacaltaya and Sierra Negra LAGO sites consists
 mainly on the measurement of the rate of signals with amplitude higher
than three different thresholds (scaler mode). The number of pulses above threshold is 
measured every 5\,ms in order to look for transient events with more than 5 sigma deviations.
The average is evaluated with 60 thousand entries (5 minutes). However, in order to match the 
efficiency of operation of the detectors and to have a calibration point,
we run also the detectors in calibration mode so  that we can get pulse height
 and integrated
charge histograms. Traces and muon decay mode are also allowed by the acquisition
 system.  We are using the electronics of the first engineering-array phase of the
Pierre Auger Observatory to readout data of the water Cherenkov detectors\cite{nim}.

\begin{figure}[!ht]
    \centering
    \includegraphics[width=0.3\textwidth]{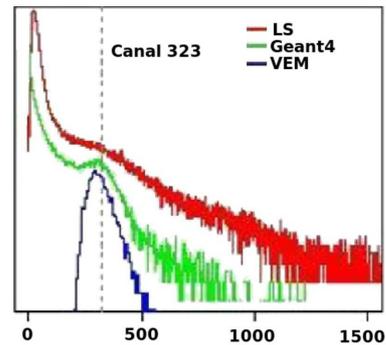}
    \caption{Real and simulated histograms of charge integrated in ADC channels 
for one of the Chacaltaya detectors.
The top curve is real data taken by the Local Station (LS). The intermediate
has been produced by Geant4 simulations.
The lower Gaussian 
distribution corresponds to vertical simulated muons.
}
    \label{simVM:chacaltaya}
  \end{figure}

We made simulations of the WCD using Geant4 and the comparison can be seen in fig.\,\ref{simVM:chacaltaya}, where a real charge histogram is compared with the simulated data,
 including a simulation corresponding to vertical muons only, used as a calibration reference. The simulations exhibit a more pronounced muon peak, probably
due to an underestimation of the flux of electromagnetic secondaries at these altitudes.
 
\begin{figure}[!ht]
    \centering
    \includegraphics[width=0.45\textwidth]{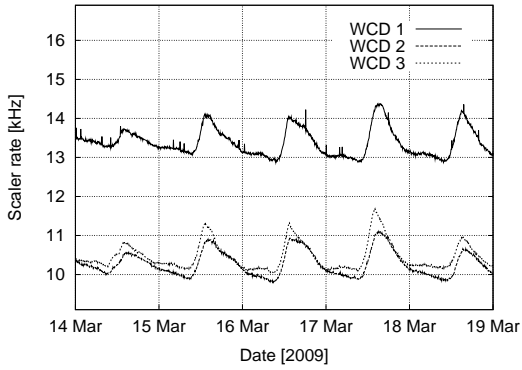}
    \caption{Five minutes average rate corresponding to the lower threshold in Sierra Negra detectors. WCD 1 has some apparent light leak, explaining its higher rate and the spikes.}
    \label{rate:lowSN}
 \end{figure}

\begin{figure}[!ht]
    \centering
    \includegraphics[width=0.45\textwidth]{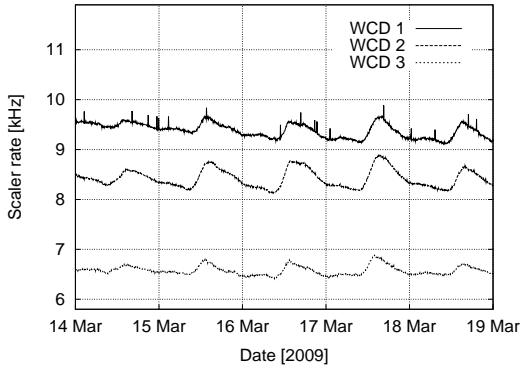}
    \caption{Five minutes average rate corresponding to the second threshold in Sierra Negra detectors. The different rates indicate calibration issues.}
    \label{rate:2dSN}
  \end{figure}

\begin{figure}[!ht]
    \centering
    \includegraphics[width=0.45\textwidth]{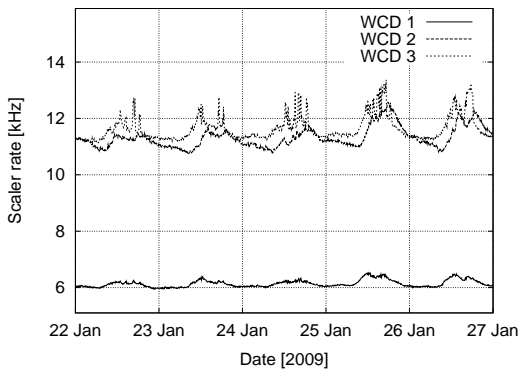}
    \caption{Five minutes average rate for the detectors at Chacaltaya corresponding to lower threshold. All three detectors are at about 3\,kHz/m$^2$ since WCD 1 is only 2\,m$^2$ while WCD 2 and 3 are 4\,m$^2$. WCD 3 exhibits some peaks
in daytime indicating light leaks.}
    \label{rate:chacaltaya}
  \end{figure}

\begin{figure}[!ht]
    \centering
    \includegraphics[width=0.45\textwidth]{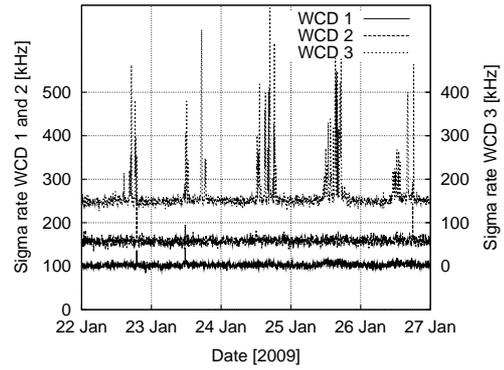}
    \caption{Stability of WCD at Chacaltaya. We plot the standard deviation
of the scaler rate every 5 minutes for the lower threshold. WCD3 clearly
exhibits a noisy behaviour during the day, proof of a light leak. Te ease the
readability, WCD 3 sigmas have been moved up 100\,Hz and the relevant scale
is indicated on the right.}
    \label{standard:deviation}
  \end{figure}

Calibration points are extracted from real data histograms by finding the
change of slope corresponding to the muons in the charge histogram.
We use them to fix the position of the three thresholds and then
 we run all the detectors in scaler mode. We show the stability of the Sierra
 Negra and Chacaltaya detectors, including rate averages for different periods of
 time and standard deviations in figs. (\ref{rate:lowSN}, \ref{rate:2dSN}, \ref{rate:chacaltaya}, \ref{standard:deviation}).

\begin{figure}[!ht]
    \centering
    \includegraphics[width=0.5\textwidth]{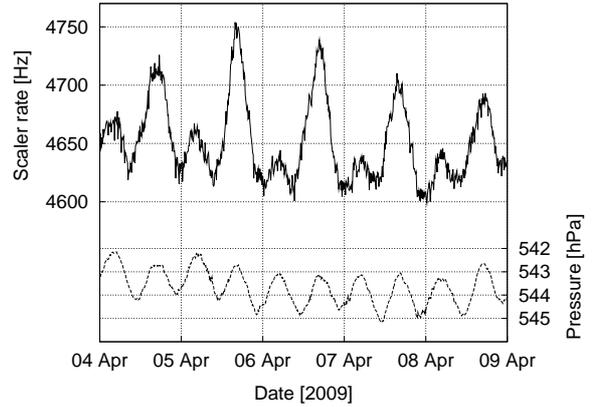}
    \caption{Second threshold scaler rate for WCD 3 at Chacaltaya and atmospheric pressure
measurements, with one data point per 10 minutes, for 5 days of April 2009. A clear
anti-correlation is found (note the scale for pressure is reversed). The rates
exhibit a stronger peak during the day (peaked at 18h local time) than at night,
indicating a likely light leak or electronic temperature-related effect.}
    \label{Chacaltaya:Pressure}
  \end{figure}

Finally we compare the second threshold scaler rate at Chacaltaya with the
atmospheric conditions in fig.\,\ref{Chacaltaya:Pressure}. The flux of
secondary particles at ground level is expected to be anti-correlated with
the atmospheric pressure, as more pressure means more absorption of the
low energy cosmic ray cascades.
The bimodal daily
variation of the pressure is present in the scaler data, but stronger 
effect is seen during day than at night. This effect could be caused either
by the effect of temperature on the electronics or some slight light leak.
The effect is under investigation.

Searches for GRBs in coincidence with satellites are reported separately in these proceedings\cite{bertou}.

   \section{New Data acquisition system} 
\begin{figure}[!ht]
    \centering
    \includegraphics[width=0.5\textwidth]{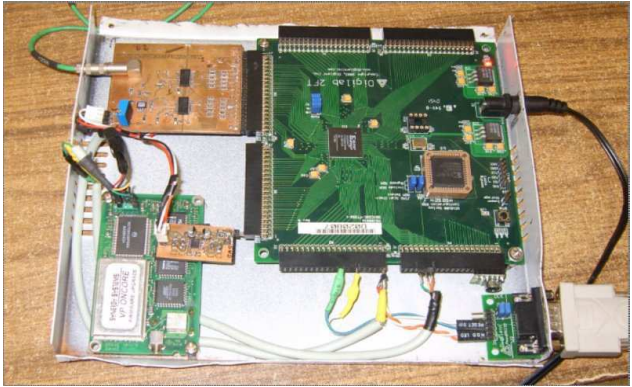}
    \caption{New electronics for LAGO. The prototype can be operated at 100 or 200\,MSPS. See text and \cite{villasen} for more details.}
    \label{new:DAQ}
  \end{figure}

Field programmable gate arrays (FPGAs) are playing an increasing role in DAQ systems in cosmic ray experiments due to their high speed and integration and their low cost and low power consumption. Modern electronics based on on-chip fast analog to digital converters (ADCs) and powerful digital signal processors (DSPs) are ideal to be the basis of custom-made DAQ systems which are much more flexible, faster and much cheaper than the traditional DAQ systems based on modular electronics\cite{villasen}. We took advantage of these recent developments, in particular in the area of very high integrated circuits in the form of ADCs and FPGAs for the design of the new system which consists of an ADC daughter board running at 200\,MSPS. Each event is tagged with precise GPS time using a GPS embedded receiver with 1 PPS (one pulse per second) synchronised with the atomic clock on the GPS satellites within a corrected uncertainty of 50\,ns (Motorola Oncore UT+ module). A pressure and Temperature sensor (HP03D) is adapted to the FPGA board (2FT Xilinx). A picture of the final setup in its RF box is visible in fig.\,\ref{new:DAQ}.

\begin{figure}[!ht]
    \centering
    \includegraphics[width=0.5\textwidth]{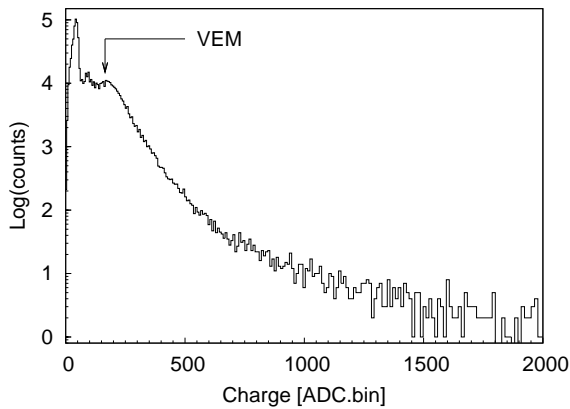}
    \caption{Calibration histogram for Sierra Negra detectors with the new DAQ, the arrow shows the point for the VEM.}
    \label{hist:calib1}
  \end{figure}

This new system provides data in the same format as the previous one. Calibration histograms can also be extracted and are shown in figure~\ref{hist:calib1} for a WCD at Sierra Negra.

As a final note, it is worth mentioning that the prototype of the HAWC experiment,
Proto-HAWC, is also located at Sierra Negra. Comparison and complementarity of LAGO and proto-HAWC could increase the sensitivity to transient events emitting high energy photons.

\section{ACKNOWLEDGEMENTS} 
The LAGO project is very thankful to the Pierre Auger collaboration for
the lending of the engineering equipment.

The author acknowledges the support of \emph{Conacyt}.

  \end{document}